\documentclass{emulateapj}
\usepackage[sort&compress]{natbib}
\usepackage{amsmath}
\usepackage{url}

\usepackage{color}
\usepackage{hyperref}
\hypersetup{colorlinks,citecolor=blue}

\newcommand{\herschel}{\textsl{Herschel}}
\newcommand{\lsim}{\mbox{$_<\atop^{\sim}$}}
\newcommand{\gsim}{\mbox{$_>\atop^{\sim}$}}

\newcommand{\lya}{\mbox{Ly--$\alpha$}}
\newcommand{\surfb}{ergs\,s$^{-1}$\,cm$^{-2}$\,arcsec$^{-2}$}

\slugcomment{{ Submitted to ApJ Letters}}

\shorttitle{WISE discovered \lya\ emitters and blobs}
\shortauthors{Bridge et~al.}

\begin{document}
\title{A New Population of High Redshift, Dusty Lyman-Alpha Emitters and Blobs Discovered by WISE}

\def\cit{1}
\def\leicester{2}
\def\ipac{3}
\def\ucla{4}
\def\goddard{5}
\def\jpl{6}
\def\sussex{7}
\def\vt{8}
\def\ucd{9}

\author{
{Carrie~R.~Bridge}\altaffilmark{\cit},
{Andrew~Blain}\altaffilmark{\leicester},
{Colin~J.K.~Borys}\altaffilmark{\ipac},
{Sara~Petty}\altaffilmark{\ucla},
{Dominic~Benford}\altaffilmark{\goddard},
{Peter~Eisenhardt}\altaffilmark{\jpl},
{Duncan~Farrah}\altaffilmark{\sussex,\vt},
{Roger~L.~Griffith}\altaffilmark{\ipac},
{Tom~Jarrett}\altaffilmark{\ipac},
{S.~Adam~Stanford}\altaffilmark{\ucd},
{Daniel~Stern}\altaffilmark{\jpl},
{Chao-Wei~Tsai}\altaffilmark{\ipac},
{Edward~L.~Wright}\altaffilmark{\ucla},
{Jingwen~Wu}\altaffilmark{\jpl}
}
\email{bridge@astro.caltech.edu}

\altaffiltext{\cit}{
California Institute of Technology
MS249-17,
Pasadena, CA 91125, USA
}

\altaffiltext{\leicester}{
Department of Physics and Astronomy,
University of Leicester,
LE1 7RH, Leicester, UK
}

\altaffiltext{\ipac}{
Infrared Processing and Analysis Center,
California Institute of Technology, MS 100-22, 
Pasadena, CA 91125, USA
}

\altaffiltext{\ucla}{
Astronomy Department, 
University of California Los Angeles,
Los Angeles, CA 90095, USA 
}

\altaffiltext{\goddard}{
NASA Goddard Space Flight Center,
Greenbelt, MD 20771, USA
}

\altaffiltext{\jpl}{
Jet Propulsion Laboratory,
California Institute of Technology,
4800 Oak Grove Dr.,
Pasadena, CA 91109, USA
}

\altaffiltext{\sussex}{
Department of Physics \& Astronomy, 
University of Sussex, 
Falmer, Brighton BN1 9QH, UK 
}

\altaffiltext{\vt}{
Department of Physics, 
Virginia Polytechnic Institute and State University, 
Blacksburg, VA 24061, USA
}

\altaffiltext{\ucd}{
Department of Physics,
University of California Davis,
One Shields Ave., Davis, CA 95616, USA
}

\begin{abstract}
We report a new technique to select $1.6\lsim z \lsim4.6$ dusty Lyman-alpha emitters (LAEs), over a third of which are `blobs' (LABs) with emission extended on scales of 30-100\,kpc.  Combining data from the NASA Wide-field Infrared Survey Explorer (WISE) mission with optical spectroscopy from the W.\,M.\,Keck telescope, we present a color criteria that yields a 78\% success rate in identifying rare, dusty LAEs of which at least 37\% are LABs.  The objects have a surface density of only $\sim0.1$\,deg$^{-2}$, making them rare enough that they have been largely missed in narrow surveys.  We measured spectroscopic redshifts for 92 of these WISE-selected, typically radio-quiet galaxies and find that the LAEs (LABs) have a median redshift of 2.3 (2.5).  The WISE photometry coupled with data from \herschel\footnote{{\it Herschel} is an ESA space observatory with science instruments provided by European-led Principal Investigator consortia and with important participation from NASA.} reveals that these galaxies have extreme far-infrared luminosities (L$_{\rm{IR}}\gsim10^{13-14}$L$_{\odot}$) and warm colors, typically larger than submillimeter-selected galaxies (SMGs) and dust-obscured galaxies (DOGs).  These traits are commonly associated with the dust being energized by intense AGN activity.  We hypothesize that the combination of spatially extended \lya, large amounts of warm IR-luminous dust,  and rarity (implying a short-lived phase) can be explained if the galaxies are undergoing strong `feedback' transforming them from an extreme dusty starburst to a QSO.

\end{abstract}


\section{Introduction}\label{s:intro}

High-redshift \lya\, emission is widely used to study star formation in galaxies \citep[e.g.][]{Cowie98,Steidel2000}.  Systems that exhibit this line are typically referred to as \lya\, emitters (LAEs); a small subset show extended emission on scales  $\gsim$30\,kpc and are considered \lya\, `blobs'  (LABs).  Among the largest coherent galactic structures known in the Universe, LABs are extremely energetic (with \lya\ luminosities of $\sim$10$^{42-44}$\,ergs\,s$^{-1}$) and have been shown to trace over-densities of galaxies at high-redshift \citep[e.g.][]{Keel1999,Francis2001, Steidel2000,Matsuda2004,Steidel2011}.   Optically-selected LABs are 100-1000 times less abundant than LAEs with only a few dozen known \citep[e.g.][]{Steidel2000,Matsuda2004}.

Resonant scattering by atomic hydrogen results in \lya\, photons being easily extinguished by dust.  Hence, blind narrow-band searches for $z\sim2-6$ LAEs and LABs have been based on the paradigm that high-redshift LAEs are generally blue galaxies, with little to no dust (A$_{V}$=0.0-2.5 mag), and moderate star formation rates \citep[10-100 M$_{\odot}$yr$^{-1}$; e.g.][]{Cowie98,Rhoads2000,Steidel2000,Matsuda2004,Gawiser2007,Gronwall2007,Nilsson2007,Nilsson2011,Finkelstein2009}.   

This suggests that dusty infrared galaxies and LAE/LAB populations, despite both peaking in activity around $z\sim2$, should largely not overlap. Indeed {\textsl{Spitzer}} observations of $\sim40$ optically selected $z=2-3.5$ LABs found that $\lsim15\%$ are luminous in the infrared \citep{Webb2009,Nilsson2009,Lai2007,Geach2005}. Furthermore, none of the optically discovered LABs \citep[e.g.][]{Steidel2000,Matsuda2004,Matsuda2011} are detected at 12 or 22\micron\ by WISE.  However, 40-50\% of the infrared 
luminous sub-millimeter galaxies (SMGs) show \lya\, emission, although with typically moderate line fluxes and rarely with 
spatially-extended  \lya\, emission \citep{Chapman2005}.  This suggests that dust and \lya\, emission are not 
in fact mutually exclusive, and points to a clumpy dust distribution that allows sight lines for \lya\, photons to escape \citep
{Neufeld1991}.  Since SMGs and related high-redshift galaxies are thought to be just one stage in the evolution of massive 
elliptical galaxies, the study of dusty galaxies with spatially extended \lya\, can thus provide a unique insight to this process. 

This paper, the first in a series, presents a new, highly efficient mid-IR color technique to select a unique population of $z\sim2-4$ dusty LAEs and LABs.   We use the redshift distribution, general spectroscopic properties, and preliminary \herschel\ observations to determine how these objects relate to other dusty galaxies at these redshifts. We assume a $\Lambda$CDM cosmology with $\Omega_{\rm M}=0.27, \Omega_\Lambda=0.73$, and H$_0$=71\,km\,s$^{-1}$\,Mpc$^{-1}$ and use Vega magnitudes.

\section{Sample Definition and Observations}
One of the primary science goals of the NASA Wide-field Infrared Survey Explorer (WISE) mission \citep{Wright2010} is to find the most luminous galaxies in the Universe.  Initial searches in WISE color space revealed a diversity of galaxy types over a wide range of redshifts \citep{Griffith2011,Wu2012}.  Deep spectroscopic follow-up of the galaxies with the reddest mid-IR colors often showed intense \lya\,emission (also see \cite{Eisenhardt2012}).  A striking feature of these WISE-selected LAEs, is that at least one-third are blobs (LABs), having spatially extended  \lya\, reaching 30-100\,kpc (see Figure~1).

\subsection{Selection Method}
\label{subsec:method}
Realizing that WISE-detected LAEs and LABs lie in a very specific region of a WISE color-magnitude diagram, we refined the selection criteria used in the initial extragalactic follow-up to focus on high-redshift LAEs/LABs. 

The resulting selection criterion, shown in Figure~2, are W2-W3 (4.5-12$\mu$m)$\ge$4.8 and SNR~$\ge$5 in W3 and W4.
We also excluded sources within 30 degrees of the Galactic center and 10 degrees of the Galactic plane to avoid AGB stars and saturation artifacts. In order to remove low-redshift ($z\lsim0.5$) star-forming galaxies with similar WISE colors, we required non-detections in the SDSS and DSS imaging ($r'\gsim22$).   Finally, each object's coadded W3 and W4 images were visually inspected to ensure the source is not spurious.    These selection criteria result in $\sim$4400 galaxies over the whole sky. 

Another WISE color selection, ``W12drop'', outlined in \cite{Eisenhardt2012} has been successful in identifying ULIRGs over a wider redshift range ($z\sim0.05-4.6$), and while there is substantial overlap in the $z\gsim2$ samples, overall the refined WISE-LAE/LAB criterion is at least 2 (2.5) times more effective at selecting high-z LAEs (LABs).  The primary differences are that the WISE-LAB criterion discussed here impose an optical magnitude limit to remove contamination from low redshift galaxies, place no flux cut on W1, and probes a factor of 1.2-2 deeper in W3 and W4 and 0.5 mags bluer in W2-W3 than the W12drop selection.

\subsection{Optical Spectroscopy}
\label{subsec:opt_spec}
To determine redshifts and discern the nature of the galaxies, we used Keck I LRIS \citep{Oke1995} over the course of six runs between July 2010 and January 2012.  The sample contains 101 objects that fulfill the WISE LAE/LAB selection, of which $\sim$45\% had already been followed-up using the W12drop criteria, and therefore did not need to be re-observed.
The majority of data were taken using the 600\,$\ell$\,mm$^{-1}$ grism in the blue arm ($\lambda_{\rm blaze}=$4000\AA; spectral resolving power $R\equiv\lambda/\Delta\lambda\sim$750), the 400\,$\ell$\,mm$^{-1}$ grating on the red arm ($\lambda_{\rm blaze}=$7800\AA; $R\sim$700), the 5600\AA\, dichroic and a 1\farcs5 wide longslit.  The targets were observed with a median seeing of 0\farcs7\, and total integration times between 20-40 minutes.  The LRIS data were reduced and flux calibrated using standard procedures.  

The optical spectroscopy reveal a wide range of characteristics, from pure starburst-dominated galaxies with narrow emission lines and several ultraviolet interstellar absorption lines comparable to Lyman break galaxies \citep{Steidel2000} to those with strong AGN components, inferred from the presence of high ionization lines such as NV, SiIV/OIV, CIV  and broad emission line profiles of several thousand km\,s$^{-1}$.   

Robust redshifts based on two or more spectral features were determined for 92 out of the 101 sources we attempted (91\% success rate), with a redshift range of $z=1.13-4.59$ (see Figure~3), while the remaining 9 showed  no continuum or spectral lines.  The excellent blue sensitivity of the LRIS detector allows \lya\, emission to be detected down to 3160\AA, corresponding to a redshift of $z\sim$1.6. Of the 92 WISE selected galaxies with a robust redshift, 89 are at a $z\ge 1.6$, and of those 79 (89\%) showed \lya\, in emission.  An additionally striking finding was that 37\% (29/79) of the WISE LAEs showed \lya\, emission in the two-dimensional spectra extended on spatial scales of $\ge$30\,kpc (i.e., LABs), with 18\% (14) having emission 40-100\,kpc in extent.  We also note that 14 have extended emission of 25-30\,kpc, which is still more than 5 times the size of the galaxies, and more than twice the seeing limitations, constituting small LABs or \lya\, clouds. 

The definition of a LAB does vary between studies.  \citet{Matsuda2004} required an isophotal area of at least 16\,arcsec$^{2}$ (or $\sim$30\,kpc in length at $z=3.1$)  and a surface brightness of $\sim$27.6 mag\,arcsec$^{-2}$ ($\sim2\times10^{-18}$\,\surfb\,), while \citet{Matsuda2011} probed  down to 1.4$\times10^{-18}$\surfb\,.  Making basic assumptions (slit width and size of extended emission) we measure a surface brightness of the $>30$\,kpc-scale blobs of 1-100$\times$10$^{-18}$\,\surfb\, which is slightly brighter than the limits used in the optical surveys.  It is therefore plausible that the true sizes of the WISE LABs are in fact larger under these standard definitions.

\subsection{Far-Infrared Photometry}
\label{subsec:far-ir}
The red WISE colors strongly argue that significant amounts of warm dust are present in these galaxies.  It is common to compare $z\sim2$ dusty galaxies by characterizing their dust luminosity and temperature.  Since their spectral energy distributions (SEDs) typically peak in the far-infrared, observations between $0.1-1$\,mm are required.  
To that end, we have undertaken a campaign with \herschel\ to study the WISE-selected LABs in five bands using the PACS \citep{pacs} and SPIRE \citep{spire} imagers.   The PACS data were obtained using the mini scan-map mode, and reached a $5\sigma$ depth of 30\,mJy at $160\mu$m.  SPIRE observations employed the small jiggle-map mode, and reached the $250\mu$m confusion limit of $\sim8\,$mJy.  The data were reduced and photometry extracted using recommended procedures with the HIPE software \citep{hipe}.  

Of the 8 objects observed so far\footnote[2]{A detailed treatment of the \herschel\ properties of WISE LABs is deferred to a separate paper when more data is collected.}, all were detected at 160\micron\, and of those, 6 are also detected at 250\micron. We fit these data using a simple SED based on a modified blackbody and in Figure 4 compare far-IR luminosity and dust temperature of the WISE LABs against SMGs and DOGs from the catalogs of \citet{Magnelli2012} and \citet{Melbourne2012}.
It is clear that the WISE LABs typically contain warmer dust, suggesting that they are being heated, at least in part, by an AGN, since starburst galaxies such as SMGs are typically colder (also see \citet{Wu2012}).  The WISE LABs are systematically more luminous than SMGs, with bolometric luminosities in excess of L$_{\rm{FIR}}>10^{13}$L$_{\odot}$, making them Hyper-Luminous Infrared Galaxies (HLIRGs). 

It might be possible to associate these high luminosities with gravitational lensing; however, none of the optical spectra demonstrate contamination from a foreground source, and the spatially extended \lya\, emission for all the sources is at the same redshift as the other spectral features.  Furthermore, adaptive-optics near-IR imaging of eight LABs with Keck's NIRC2 camera \citep{Bridge2012,Eisenhardt2012}  reveal no potential lens (nearby galaxies or clusters) or shearing of the source.

\begin{figure}
\label{fig:spectra}
\begin{center}
\epsscale{1.05}
\rotatebox{0}{\hspace{-0.2cm}
\plotone{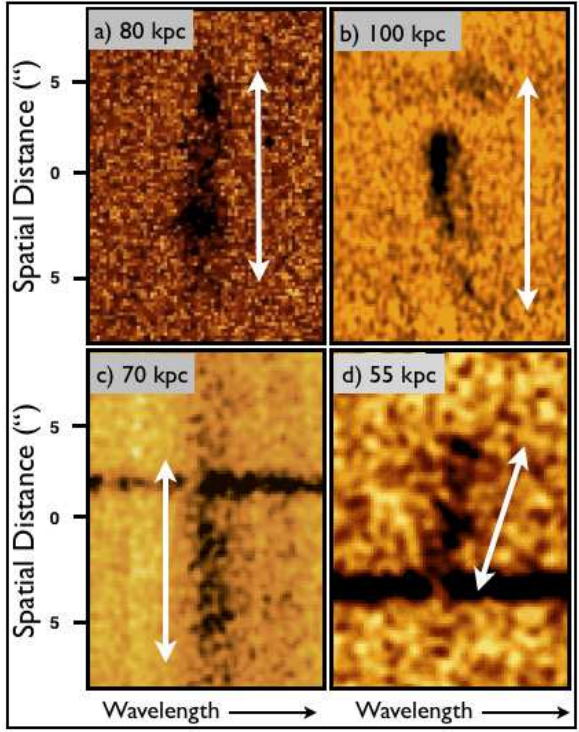}}
\caption{Two-dimensional LRIS spectra for four of the WISE LABs in the sample.  The spatial extent of the \lya\, emission is noted in the corner of each image.  These examples are representative of the asymmetric, spatially extended line profile, and clumpy nature of the \lya\, emission found in these galaxies.}
\end{center}
\end{figure}

\begin{figure}
\label{fig:colorsel}
\begin{center}
\epsscale{1.35}
\rotatebox{0}{\hspace{-0.85cm}
\plotone{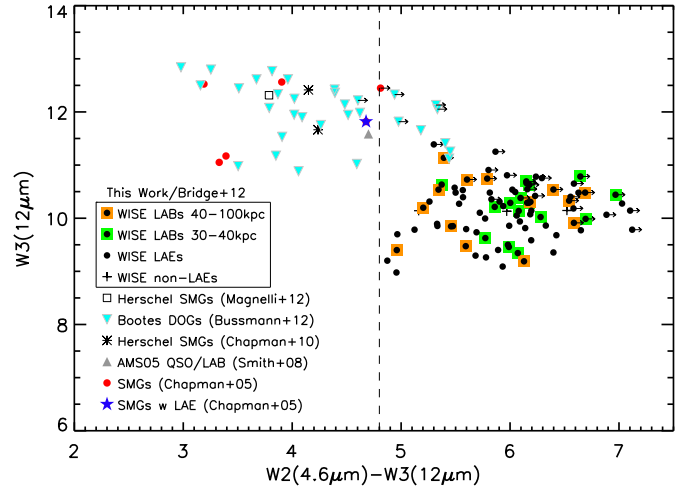}}
\caption{ WISE W3(12\micron\,) vs. WISE W2-W3 (4.5-12$\mu$m) color for all sources in the sample with spectroscopic redshifts between $1.6\le z\le 4.5$.  WISE LAEs (black circles), 30-40\,kpc LABs (green squares), and 40-100kpc LABs (orange squares) are plotted.
For comparison, $z=2-3$ WISE detected SMGs and DOGs are highlighted.  The WISE LAEs (LABs) exhibit unique, redder mid-IR colors (implying hotter dust temperatures) than other dusty $z\sim2$ populations.W2-W3 color selection criteria has a 78\% ($>$29\%) success rate in identifying $z>2$ dusty LAEs (LABs) (dashed line).}
\end{center}
\end{figure}

\begin{figure}
\label{fig:redist}
\begin{center}
\epsscale{1.3}
\rotatebox{0}{\hspace{-0.85cm}
\plotone{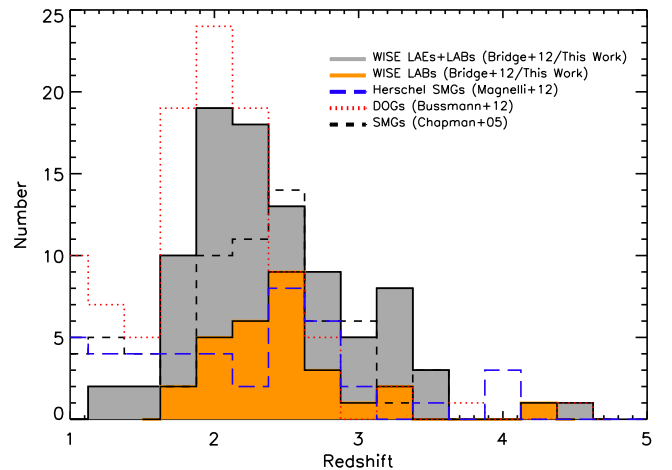}}
\caption{Spectroscopic redshift distribution for the full sample (LAEs+LABs; filled grey), and WISE LABs (filled orange).  For comparison we plot the redshift distribution of other well-studied dusty populations: SMGs (black dashed; \cite{Chapman2005}), Herschel SMGs (blue long dashed; \cite{Magnelli2012}), and DOGs (red dotted; \cite{Bussmann2012}).   }
\end{center}
\end{figure}

\begin{figure}
\begin{center}
\epsscale{1.15}
\rotatebox{0}{\hspace{-0.45cm}
\plotone{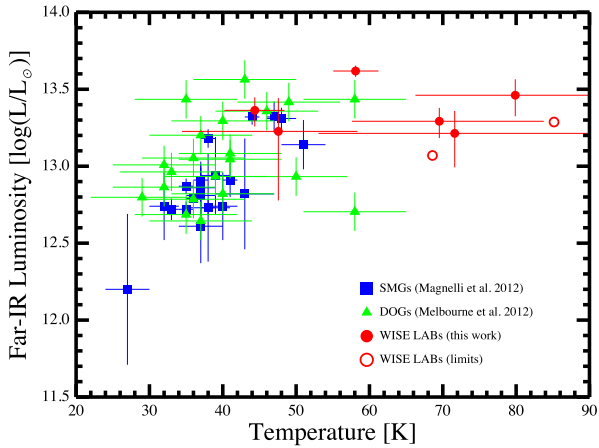}}
\caption{Estimated dust properties based on \herschel\, observations.  Objects between $2.0\leq z \leq3.5$ extracted from catalogs of DOGs and unlensed SMGs are plotted against the eight observed WISE LABs over the same redshift range.   In all cases, fits are to a single temperature modified blackbody.  WISE objects without a detection at 250$\mu$m are plotted as open circles, and represent lower limits to both the temperature and luminosity.  Error bars in all cases are based on fit results only and do not account for other systematic effects.}
\end{center}
\label{fig:herschel}
\end{figure}

\section{Discussion}
\label{sec:discussion}
One of the main features that sets the WISE LAEs apart from other high-redshift galaxies is the unprecedented high fraction that exhibit spatially extended \lya\, emission, despite having large amounts of dust.    We cross matched the WISE catalog with all previously known optically selected LABs in the literature and found that {\it none} are detected at 12 or 22$\mu$m above a SNR of 5, including the ones known to host ULIRGs \citep{Geach2005,Prescott2009}.  Despite being a factor of 10-1000 times more luminous in the mid-far-IR, the WISE LABS have \lya\, luminosities (10$^{42-44}$\,ergs\,s$^{-1}$), rest-frame equivalent widths (6-300\AA\,), and blobs sizes similar to optically discovered LABs.    With a surface density of 0.03 per square degree however, the WISE LABs are thousands of times more rare.

\subsection{WISE LABs and Other Dusty High-Redshift Populations}
\label{subsec:compare}
Though the IR luminosity and redshift distribution of the WISE LAEs/LABs are similar to other populations of IR-luminous galaxies (Figure~3), they must differ in some way given the large fraction with extended \lya\, emission.  We now compare the WISE LAEs/LABs to other high-redshift dusty populations and also note how the various selections overlap.  A more in-depth comparison will be presented in the upcoming catalog paper \citep{Bridge2012}.

Blind sub-millimeter surveys have been very successful at finding high-redshift, dust-obscured galaxies with far-IR luminosities in excess of $10^{12}$L$_{\odot}$ (i.e. SMGs).  However, these long wavelength observations ($\gsim500\,\mu$m) select systems with cooler dust and less extreme mid-IR colors than seen in our WISE sample (Figure 2 \& 4).  Generally, SMGs are powered by star formation rather than Compton-thin AGN \citep{Alexander2005}.  Based on the limited \herschel\ observations conducted so far, all the WISE LABs would also be classified as SMGs (Figure~4). However only 1-2\% of known SMGs fulfil the WISE LAE/LAB criteria, and there are only a few rare SMGs with extended \lya\, emission.  

\citet{Dey2008} have successfully used a mid-IR to optical flux ratio ($F_{24\micron}/F_{R}>$1000 and  $F_{24\micron}>0.3$mJy)  to select a population of $z\sim2$ dust-obscured galaxies \citep[DOGS;][]{Houck2005,Desai2009,Bussmann2012}.  By probing blueward of the SED peak, this technique recovers a mix  of AGN and starburst dominated galaxies.  Although all the WISE LAEs/LABs fulfil this criteria (using $r'-$W4), only $\lsim8\%$ of the DOGs in \citet{Bussmann2012} fulfill the WISE selection presented here.   WISE LAEs/LABs are also significantly brighter in the mid-far-IR ($>$5 times).  Finally, the DOGs criterion does not specifically select sources with a steeply rising mid-IR SED (Fig.~2), that we have shown are generally the objects that exhibit extreme \lya\, properties. 

Our sources are typically undetected or only faintly detected in the FIRST ($<$1mJy) or NVSS radio surveys, differentiating them from the well-studied high-redshift radio galaxies \citep{Reuland2003} and QSOs which can also exhibit large \lya\ halos \citep[e.g.][]{McCarthy1987,Smith2008}.   

In short, though surveys of DOGs/SMGs will also find WISE LAEs/LABs,  with a projected source density of roughly 1 per 30 deg$^{2}$, only large-area/all-sky surveys are able to find a statistically significant sample.

\subsection{Spectral Properties of WISE LABs}
\label{subsec:specprop}
The cause of the LAB phenomena in optically selected samples is still debated \citep[e.g.][] {Prescott2009,Colbert2011}, but given that all WISE LABs are associated with a luminous IR galaxy while less than 15\% of optical LABs are \citep[e.g.][]{Webb2009,Nilsson2009}, the dominant powering mechanisms behind these two species of LABs are likely different. For clues we investigate the optical spectra.

The high-ionization and broad emission lines seen in $\gsim$70\% of optical spectra suggest that the majority, if not all, of the WISE LAEs/LABs host a dust-obscured AGN, which is also consistent with the \herschel\ and WISE colors. The \lya\, emission is spatially asymmetric and clumpy,  and is often offset from the systemic redshift of the galaxy, implying large-scale flows.  Furthermore, many galaxies in the sample demonstrate a \lya\ velocity structure redshifted up to several thousand km\,s$^{-1}$ as a function of distance from the central component (Figure~1d).  The line profiles are diverse,  ranging from traditional P-Cygni profiles characteristic of Wolf-Rayet and O-star winds to \lya\, emission peaking on the blue side of the \lya\, profile, suggesting possible inflows \citep{Dijkstra2009,Barnes2011}.  

Multiple slit orientations were used to probe the spatial line morphology for four of the LABs.  A slit position angle rotated by 45 and 90 degrees of the original angle often showed little or no extended emission, implying an asymmetric and filamentary line morphology.  This result suggests that the quoted 
efficiency of the color selection in identifying LABs is a lower limit.  Indeed, using a simple model of the extended emission size and geometry, and assuming that all LAEs in our sample are LABs,  the 37\% detection rate is consistent with a random slit orientation.

\subsection{WISE LABs: Feedback Caught in the Act?}
\label{subsec:feedback}

The correlation between the masses of supermassive black holes (SMBH) at the center of nearby ellipticals and their bulge stellar velocity dispersion remains striking in the study of galaxy evolution  \citep[e.g.][]{Ferrarese2000}.  The general paradigm that ties these passive galaxies with their high-redshift progenitors is a process of merger-induced star formation which also fuels the SMBH \citep[e.g.][]{Sanders1996,Hopkins2006}. The system appears at one point as a heavily obscured star forming galaxy, and later, as the SMBH accretes at a higher rate, a galaxy with an active nucleus.  AGN- and starburst-induced winds eventually become strong enough to expel the obscuring gas and dust, briefly revealing an optical quasar.  This short-lived `feedback' process is thought to quench both star formation and AGN activity, leading ultimately to a passive, red galaxy spheroid \citep[e.g.][]{Hopkins2006,Farrah2012}.   Observational evidence for this process has been challenging to obtain, but models provide a general picture of the properties of these systems. 

It is well established that the peak of this feedback activity occurs near  $z\sim2$ \citep[e.g.][]{DiMatteo2005}, and that a multitude of galaxy types are associated with the process.  SMGs are thought to be the IR-luminous starbursting pre-cursors to systems that will evolve into a massive elliptical, and have been well characterized in a number of studies \citep[e.g.][]{Chapman2005,Borys2005,Magnelli2012}.

Using a 3D radiative equilibrium code, \citet{Chakrabarti2007} showed that AGN feedback is particularly effective at dispersing gas and dust.  
They predict that for a brief time ($<40$Myr) while the AGN injects energy into the galaxy and surrounding IGM at its maximal rate, the dust is hotter than during the SMG starburst phase.  Strikingly, all the WISE LABs observed with \herschel\, thus far show these warm far-IR colors consistent with being in a short-lived AGN feedback phase.  The outflow of material would not necessarily be symmetrical,  and \lya\ could appear filamentary.

The WISE LABs which demonstrate warm, IR-luminous dust and directional \lya\, emission,  share these general properties.  The surface density of LABs is generally consistent with predicted numbers of HLIRGs from models \citep[e.g.][]{Bethermin2011} coupled with timescale arguments for the feedback process and length of the SMG phase.  However we caution that this is not a particularly strong argument given the uncertainty in the models. Nevertheless, the accumulated evidence suggesting that the WISE LABs are undergoing extremely powerful feedback is intriguing, and we will expand on it in future papers.

\section{Conclusions and Future Work}
\label{sec:conclusions}

This letter presents a new WISE color criterion with a 78$\%$ success rate in selecting $z=1.6-4.6$ dusty, mid-IR bright,  \lya\, emitters, of which at least 37$\%$, are blobs.  This new population of galaxies, largely missed in narrower surveys,  have a redshift distribution that peaks at $z\sim$2.3, total IR luminosities on average brighter than SMGs and other high-redshift dusty galaxies,  and a density of $\sim$ 0.1 deg$^{-2}$. This is the first systematic search technique to highlight dusty LAEs/LABs, and unlike optical narrow-band searches covers a large redshift range, the whole sky, and without contamination by [OII] interlopers, making it well suited to providing targets for future large spectroscopic surveys. 

We speculate that these galaxies are in a short-lived transition between a dusty starburst and an optical QSO driven by the central AGN.  If true, these systems offer a unique opportunity to investigate AGN feedback and how it can effect not only the galaxy but also the surrounding intergalactic medium.  A full census of the WISE LAEs/LABs, optical spectroscopy, mid-far-IR properties, and near-IR morphologies will be presented in a series of forthcoming papers.

\begin{acknowledgments} 
\textit{Facilities:}
\facility{WISE, Keck (LRIS),\herschel\, (PACS, SPIRE)}
\bigskip

Acknowledgements: This publication makes use of data products from the Wide-field Infrared Survey Explorer, which is a joint project of the University of California, Los Angeles, and the Jet Propulsion Laboratory/California Institute of Technology, funded by the National Aeronautics and Space Administration.

Some of the data presented herein were obtained at the W.M. Keck Observatory, which is operated as a scientific partnership among the California Institute of Technology, the University of California and the National Aeronautics and Space Administration. The Observatory was made possible by the generous financial support of the W.M. Keck Foundation. 

The authors wish to recognize and acknowledge the very significant cultural role and reverence that the summit of Mauna Kea has always had within the indigenous Hawaiian community.  We are most fortunate to have the opportunity to conduct observations from this mountain.

\end{acknowledgments}

\bibliographystyle{apj}

\end{document}